\documentclass[aps,preprint,prl,superscriptaddress]{revtex4}

\usepackage{graphicx}
\usepackage{amsmath}
\usepackage{amssymb}
\usepackage{units}
\usepackage{color}

\usepackage{setspace}

\begin{document}

\title{Enhanced electron-phonon coupling in graphene with periodically distorted lattice}

\author{E. Pomarico}
\email{enrico.pomarico@epfl.ch}
\affiliation{Max Planck Institute for the Structure and Dynamics of Matter, Center for Free Electron Laser Science, Hamburg, Germany}
\affiliation{Laboratory for Ultrafast Microscopy and Electron Scattering, Institute of Physics, Ecole Polytechnique F{\'e}d{\'e}rale de Lausanne (EPFL), Lausanne, Switzerland.}
\author{M. Mitrano}
\affiliation{Max Planck Institute for the Structure and Dynamics of Matter, Center for Free Electron Laser Science, Hamburg, Germany}
\affiliation{Department of Physics and Frederick Seitz Materials Research Laboratory, University of Illinois, Urbana, Illinois 61801, USA}
\author{H. Bromberger}
\affiliation{Max Planck Institute for the Structure and Dynamics of Matter, Center for Free Electron Laser Science, Hamburg, Germany}
\author{M. A. Sentef}
\affiliation{Max Planck Institute for the Structure and Dynamics of Matter, Center for Free Electron Laser Science, Hamburg, Germany}
\author{A. Al-Temimy}
\author{C. Coletti}
\affiliation{Center for Nanotechnology @ NEST, Istituto Italiano di Tecnologia, Pisa, Italy}
\author{A. St{\"o}hr}
\author{S. Link}
\author{U. Starke}
\affiliation{Max Planck Institute for Solid State Research, Stuttgart, Germany}
\author{C. Cacho}
\author{R. Chapman}
\author{E. Springate}
\affiliation{Central Laser Facility, STFC Rutherford Appleton Laboratory, Harwell, United Kingdom}
\author{A. Cavalleri}
\affiliation{Max Planck Institute for the Structure and Dynamics of Matter, Center for Free Electron Laser Science, Hamburg, Germany}
\affiliation{Department of Physics, Clarendon Laboratory, University of Oxford, Oxford, United Kingdom}
\author{I. Gierz}
\email{Isabella.Gierz@mpsd.mpg.de}
\affiliation{Max Planck Institute for the Structure and Dynamics of Matter, Center for Free Electron Laser Science, Hamburg, Germany}

\date{\today}

\begin{abstract}
Electron-phonon coupling directly determines the stability of cooperative order in solids, including superconductivity, charge and spin density waves. Therefore, the ability to enhance or reduce electron-phonon coupling by optical driving may open up new possibilities to steer materials' functionalities, potentially at high speeds. Here we explore the response of bilayer graphene to dynamical modulation of the lattice, achieved by driving optically-active in-plane bond stretching vibrations with femtosecond mid-infrared pulses. The driven state is studied by two different ultrafast spectroscopic techniques. Firstly, TeraHertz time-domain spectroscopy reveals that the Drude scattering rate decreases upon driving. Secondly, the relaxation rate of hot quasi-particles, as measured by time- and angle-resolved photoemission spectroscopy, increases. These two independent observations are quantitatively consistent with one another and can be explained by a transient three-fold enhancement of the electron-phonon coupling constant. The findings reported here provide useful perspective for related experiments, which reported the enhancement of superconductivity in alkali-doped fullerites when a similar phonon mode was driven.  
\end{abstract}

\maketitle

\section{introduction}

Periodically driven systems can exhibit entirely different properties compared to those at equilibrium, as the modulation can give rise to new effective Hamiltonians \cite{PolkovnikovRevModPhys2011}. Classical examples range from dynamically stabilized inverted pendula \cite{KapitzaJETP1951} to ion traps \cite{PaulNobel1989}.

In low dimensional solids, periodic electromagnetic fields have been predicted \cite{LindnerNatPhys2011, SentefNComms2015} and shown \cite{WangScience2013, MahmoodNatPhys2016} to modify electronic structure and topology. Also, important changes in the functional properties of complex solids have been demonstrated when the lattice was directly driven at mid-infrared frequencies. Lattice excitation was found to induce new crystal structures \cite{FörstNatPhys2011, FörstSSC2013, MankowskyNature2014, SubediPRB2014, MankowskyPRB2015}, and to lead to a number of unexpected phenomena including the enhancement of superconductivity \cite{FaustiScience2011, KaiserPRB2014, HuNatMater2014, HuntPRB2015, MitranoNature2016}, photo-stimulated insulator-to-metal transitions \cite{RiniNature2007, CavigliaPRL2012, HuPRB2016}, and the melting of magnetic \cite{FörstPRB2011, CavigliaPRB2013} or orbital order \cite{TobeyPRL2008}.

\begin{figure}
	\center
  \includegraphics[width = 0.5\columnwidth]{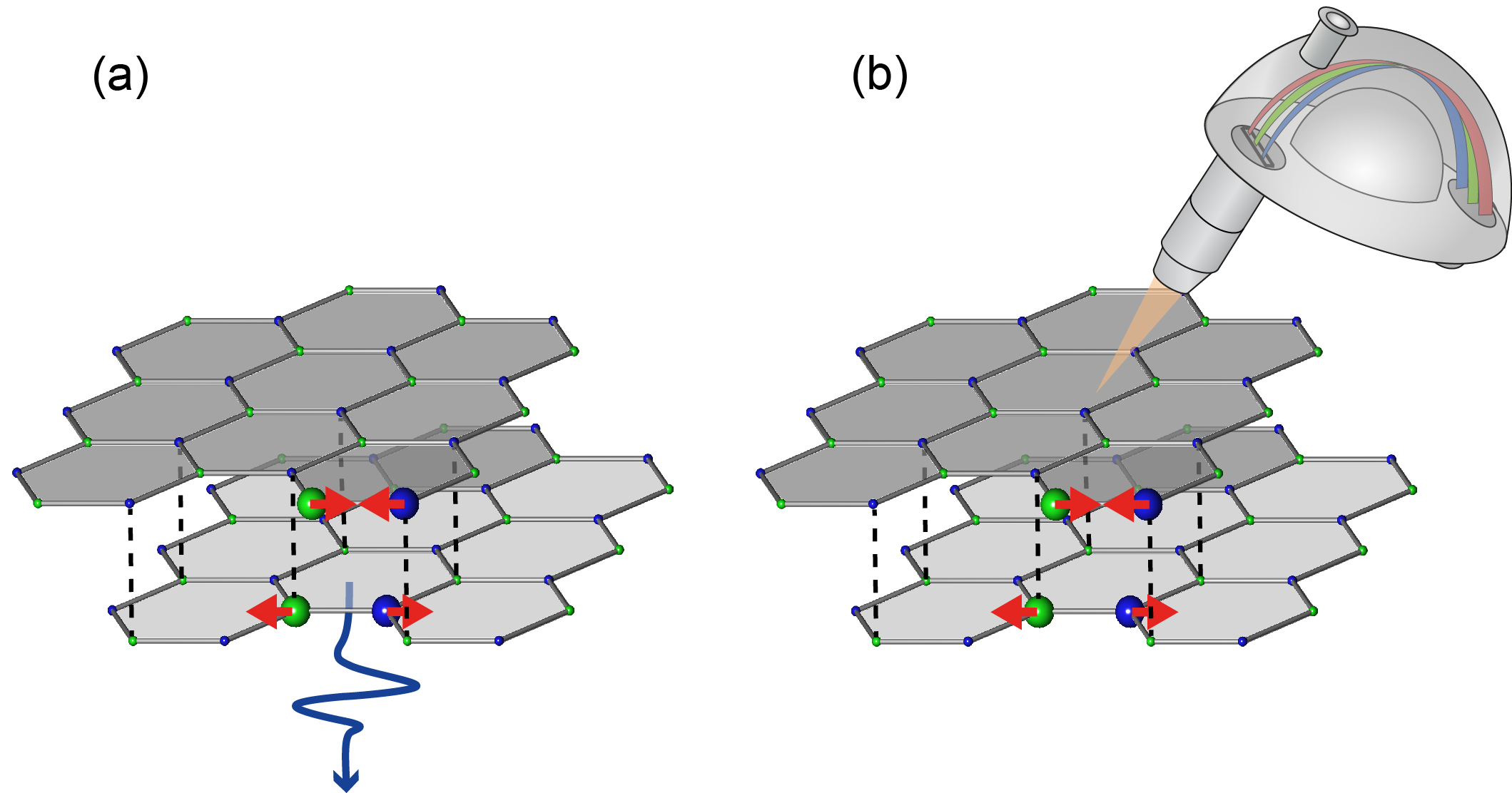}
  \caption{The effects of resonant excitation of the in-plane E$_{\text{1u}}$ mode in bilayer graphene (the atomic motions are indicated by red arrows) are investigated with TeraHertz time-domain spectroscopy (a) and time-and angle-resolved photoemission spectroscopy (b).}
  \label{figure1}
\end{figure}

Here, we set out to measure the electron-phonon interaction in graphene in the presence of a dynamical modulation of the lattice. We excited epitaxial bilayer graphene grown on SiC(0001) \cite{RiedlPRL2009, BiancoAPL2015}, both in and out of resonance with the in-plane bond-stretching E$_{\text{1u}}$ phonon at 200\,meV (6.3\,$\mu$m wavelength). The electronic properties of the driven state were then measured with both TeraHertz time-domain spectroscopy (THz TDS, Fig. \ref{figure1}a) and time- and angle-resolved photoemission spectroscopy (tr-ARPES, Fig. \ref{figure1}b). We find a three-fold enhancement of the electron-phonon coupling constant $\lambda_{\text{e-ph}}$ when the pump pulse is tuned to the E$_{\text{1u}}$ phonon resonance.

\section{methods}

\subsection{sample growth}

The bilayer graphene samples for the time-resolved optical and ARPES experiments were prepared according to two slightly different recipes. The samples for the optical experiments were obtained by thermal decomposition of the silicon face of silicon carbide (SiC) \cite{BiancoAPL2015}. In this case, the interface between bilayer graphene and the SiC substrate is formed by a carbon buffer layer. This results in n-doped samples with the chemical potential at $+0.3$\,eV inside the conduction band. For the time-resolved ARPES experiments, quasi free-standing bilayer samples have been used \cite{RiedlPRL2009}. These samples are obtained by forming a graphene monolayer on top of a carbon buffer layer by thermal decomposition of the substrate. The buffer layer is then turned into a second graphene layer by hydrogen intercalation. In this case, bilayer graphene rests on a hydrogen-terminated SiC substrate, resulting in p-doped samples with the chemical potential at $-0.12$\,eV inside the valence band. In order to enable THz TDS experiments in transmission geometry the amorphous carbon layers on the backside of the SiC substrate were removed by oxygen plasma etching.

\subsection{time-resolved measurements}

Both THz TDS and tr-ARPES xperiments made use of femtosecond pulses from a Titanium:Sapphire amplifier at a repetition rate of 1\,kHz. The mid-infrared pump pulses were generated with an optical parametric amplifier with difference frequency mixing. All experiments were performed with a pump fluence of 0.3\,mJ/cm$^2$. THz probe pulses at frequencies between 0.5 and 2.5\,THz were generated in a GaAs photoconductive antenna, and detected in the time domain by electro-optic sampling after transmission through the sample. Extreme ultra-violet pulses for the tr-ARPES experiment were generated by high harmonics generation in argon. A single harmonic at $\hbar\omega_{\text{probe}}=30$\,eV was selected with a time-preserving grating monochromator \cite{FrasettoOE2011}. The ejected photoelectrons were dispersed according to their emission angle and kinetic energy with a hemispherical analyzer and counted on a two-dimensional microchannel plate detector.

\section{results}

\subsection{THz TDS data}

In order to obtain the equilibrium opical properties of our sample we measured the THz field transmitted through bilayer graphene on top of the SiC substrate, $E$, as well as through the bare substrate, $E_{\text{ref}}$. The complex transmittance of bilayer graphene is then calculated via $T=E/E_{\text{ref}}$. For a thin film on an insulating substrate, the transmittance is related to the complex optical conductivity $\sigma = \sigma_1 +i\sigma_2$ via $T=\frac{1+N}{1+N+Z_0 \sigma d}$ \cite{AverittJPCM2002}, where $N=2.55$ is the refractive index of the SiC substrate, $Z_0=377\Omega$ is the free space impedance, and $d$ is the thickness of the thin film (the graphene bilayer). Throughout this work we use $\sigma\times d$ in units of $\Omega^{-1}$. The equilibrium optical properties in the THz spectral range are well described by the Drude model \cite{WangScience2008, LiNatPhys2008, HongPRB2011}. The real part of the optical conductivity is given by $\sigma_1=\frac{\sigma_{\text{DC}}}{1+(\omega\tau_{\text{D}})^2}$ with a zero-frequency term $\sigma_{\text{DC}}=\frac{Ne^2\tau_{\text{D}}}{m^*}$ \cite{DresselBook2002}. Using this to fit the measured $\sigma_1$, we obtain a DC conductivity of $\sigma_{\text{DC}}^{\text{eq}} =(2.3 \pm 0.2$)$10^{-3} \Omega^{-1}$ and a scattering time of $\tau_D^{\text{eq}}= (95 \pm 7)$\,fs. 

\begin{figure}
	\center
  \includegraphics[width = 0.5\columnwidth]{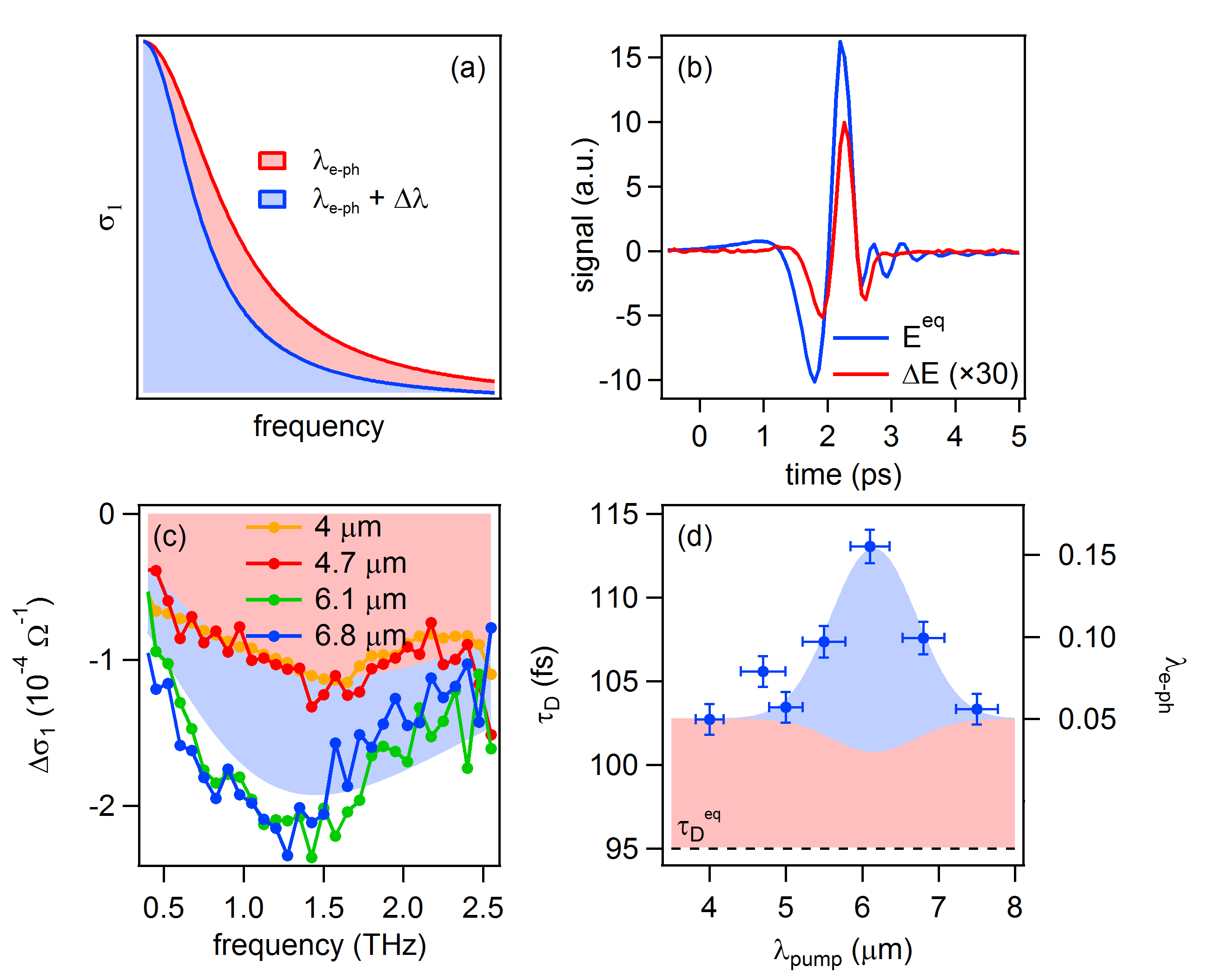}
  \caption{(a) An increase of the electron-phonon coupling constant $\lambda_{\text{e-ph}}$ is predicted to enhance the scattering time and thus reduce the width of the Drude peak in the real part of the optical conductivity. (b) Transmitted electric THz field without pump (blue) and pump-induced changes of the field at the peak of the pump-probe signal for $\lambda_{\text{pump}}=6.1$\,$\mu$m. (c) Pump-induced changes of the real part of the optical conductivity as a function of energy for different pump wavelengths in the mid-infrared together with Drude fits. (d) Drude scattering time (left axis) and electron phonon-coupling constant (right axis) as a function of pump wavelength. The red(blue)-shaded area indicates the increase in $\tau_{\text{D}}$ due to heating (an enhanced $\lambda_{\text{e-ph}}$). The black dashed line indicates the equilibrium value of $\tau_{\text{D}}$.}
  \label{figure2}
\end{figure}

Note that electron-phonon coupling $\lambda_{\text{e-ph}}$ is expected to increase both the Drude scattering time $\tau_{\text{D}}$ and the effective electron mass $m^*$ by a factor of $(1+\lambda_{\text{e-ph}})$ \cite{DresselBook2002}. Hence, a change in electron-phonon scattering changes the width of the Drude peak but leaves $\sigma_{\text{DC}}$ unaffected, as illustrated in Fig. \ref{figure2}a. 

In Fig. \ref{figure2}b we show the THz pulse after transmission through bilayer graphene (blue) and the pump-induced changes of the pulse for excitation at $\lambda_{\text{pump}}=6.1$\,$\mu$m (red), as measured in the time domain. After Fourier transformation, we directly obtained the pump-induced changes of the optical conductivity from the relative transmission changes $\Delta E/E$. In the limit of a film much thinner than the penetration depth of the pump pulse, the changes in conductivity can be expressed as $\Delta\sigma=-\frac{1+N}{Z_0}\frac{\Delta E}{E}$ \cite{JohnstonBook2012}.
 
The pump-induced changes of the real part of the optical conductivity $\Delta\sigma_1$ are shown in Fig. \ref{figure2}c. Fitting the transient data with the Drude model revealed that $\sigma_{\text{DC}}$ was slightly reduced after excitation by up to $\sim$2\% compared to the equilibrium value with no systematic pump wavelength dependence (not shown here). On the other hand, $\tau_{\text{D}}$ exhibited a pronounced increase when the pump was tuned to the E$_{\text{1u}}$ phonon resonance (Fig. \ref{figure2}d). This observation is consistent with a scenario in which electronic heating (red-shaded area in Fig. \ref{figure2}d) dominates the response for non-resonant excitation, with an additional increase in electron-phonon coupling at resonance to the phonon (blue-shaded area in Fig. \ref{figure2}d). Starting from the previously reported equilibrium value of the electron-phonon coupling $\lambda_{\text{e-ph}}^{\text{eq}}=0.05$ \cite{ParkNanoLett2008, SiegelNJP2012, JohannsenPRL2013}, we estimate a three-fold increase of $\lambda_{\text{e-ph}}$ for the periodically distorted graphene lattice (right axis in Fig. \ref{figure2}d) using $$\lambda_{\text{e-ph}}(\lambda_{\text{pump}})=\left(1+\lambda_{\text{e-ph}}^{\text{eq}}\right)\frac{\tau_D(\lambda_{\text{pump}})}{\tau_D^{\text{off}}}-1,$$ where $\tau_D^{\text{off}}=103$\,fs is the transient Drude scattering time measured far away from the phonon resonance. 

\subsection{tr-ARPES data}

\begin{figure}
	\center
  \includegraphics[width = 1\columnwidth]{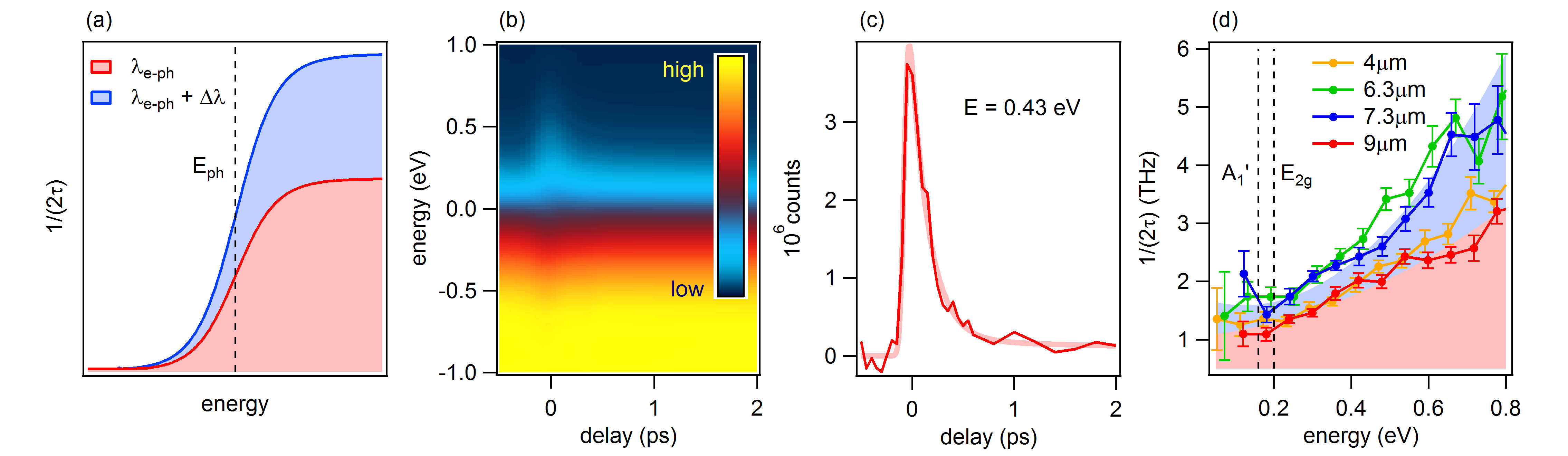}
  \caption{(a) Expected increase of the energy-resolved relaxation rate due to a pump-induced enhancement of $\lambda_{\text{e-ph}}$. (b) Momentum-integrated photocurrent measured along the $\Gamma$K-direction in the vicinity of the K-point as a function of pump-probe delay after excitation at $\lambda_{\text{pump}}=6.3$\,$\mu$m. These experiments were performed at 30\,K with energy and temporal resolutions of 600\,meV and 100\,fs, respectively. (c) Lineouts at constant energy (here $E=0.43$\,eV) are fitted with an exponential decay to obtain the energy-dependent relaxation time $\tau$. (d) Relaxation rate $1/(2\tau)$ as a function of energy for different pump wavelengths in the mid-infrared. For $\lambda_{\text{pump}}=6.3$\,$\mu$m and 7.3\,$\mu$m (at resonance with the E$_{\text{1u}}$ phonon) the energy-integrated relaxation rate is enhanced, indicating an increase of the electron-phonon coupling constant $\lambda_{\text{e-ph}}$.}
  \label{figure3}
\end{figure}

This interpretation is corroborated by tr-ARPES measurements. Note that for non-equilibrium states, the conventional line width analysis used to obtain $\lambda_{\text{e-ph}}$ in equilibrium \cite{BostwickNatPhys2007}, cannot be used \cite{KemperPRB2014}. Rather, an enhancement of $\lambda_{\text{e-ph}}$ appears as an increase of the relaxation rate of hot carriers at all energies \cite{KemperPRB2014, SobotaJES2014, SentefPRX2013} as illustrated in Fig. \ref{figure3}a.

Fig. \ref{figure3}b shows the momentum-integrated photocurrent measured along the $\Gamma$K-direction in the vicinity of the K-point as a function of energy and pump-probe time delay for $\lambda_{\text{pump}}=6.3$\,$\mu$m. Exponential fits of transient lineouts at constant energy (Fig. \ref{figure3}c) yielded the energy-dependent relaxation rates $1/(2\tau)$ shown in Fig. \ref{figure3}d. The same measurement and analysis was repeated for a similar range of pump wavelengths as in the optical experiment in Fig. \ref{figure2}. For on-resonance excitation ($\lambda_{\text{pump}}=6.3$\,$\mu$m and 7.3\,$\mu$m) we observed systematically higher relaxation rates than for off-resonance excitation ($\lambda_{\text{pump}}=4$\,$\mu$m and 9\,$\mu$m). In agreement with Fig.  \ref{figure2}d, this observation is also suggestive of a transiently enhanced electron-phonon coupling constant $\lambda_{\text{e-ph}}$, which occurs only for excitation at resonance to the E$_{\text{1u}}$ phonon.

\begin{figure}
	\center
  \includegraphics[width = 0.5\columnwidth]{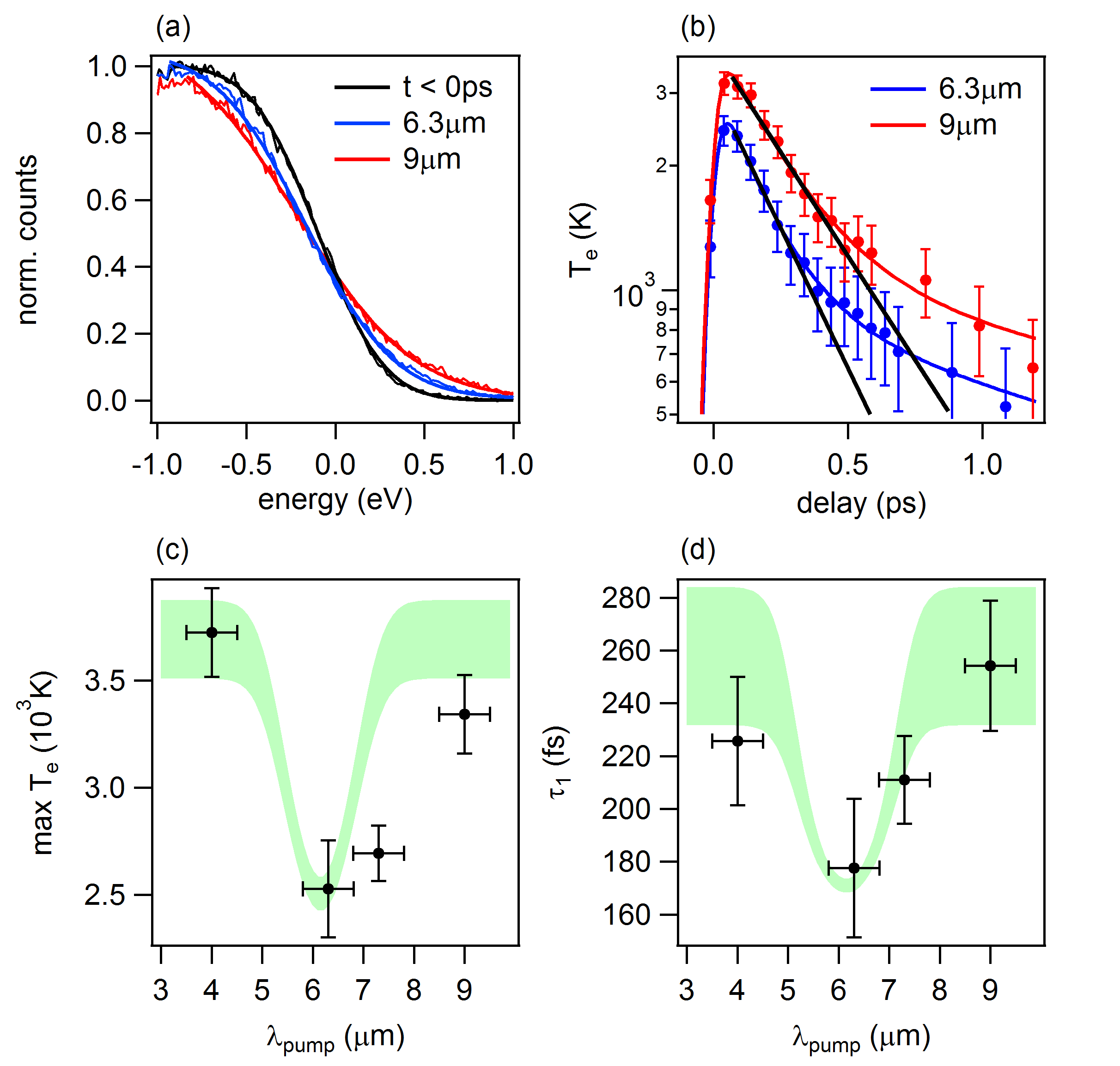}
  \caption{(a) Momentum-integrated photocurrent as a function of energy at negative delay (black) and at the peak of the pump-probe signal for $\lambda_{\text{pump}}=6.3$\,$\mu$m (on-resonance, blue) and $\lambda_{\text{pump}}=9$\,$\mu$m (off-resonance, red) together with Fermi-Dirac fits. (b) Electronic temperature as a function of pump-probe delay for $\lambda_{\text{pump}}=6.3$\,$\mu$m (blue) and $\lambda_{\text{pump}}=9$\,$\mu$m (red) together with double-exponential fits. Black lines illustrate the difference in the fast time constant between on- and off-resonance excitation. Electronic peak temperature (c) and fast relaxation time (d) as a function of pump wavelength (data points) together with a simulation based on a two-temperature model (green lines). The width of the green lines represents the experimental uncertainty in $\lambda_{\text{e-ph}}$. }
  \label{figure4}
\end{figure}

The interpretation above is significantly strengthened by a quantitative comparison of the two complementary approaches. From the THz TDS data, we have estimated a three-fold increase in $\lambda_{\text{e-ph}}$. Such enhancement in $\lambda_{\text{e-ph}}$ is also expected to affect both the peak electronic temperature and the cooling rate of the hot electron population \cite{GierzPRL2015}. Figure \ref{figure4}a shows measured electron distributions, obtained by integrating the photocurrent over momentum, together with Fermi-Dirac fits. The black curve was measured at negative delay. The blue and red curves were measured at the peak of the pump-probe signal for excitation on- and off-resonance with the E$_{\text{1u}}$ phonon, respectively. The corresponding temporal evolution of the electronic temperature is shown in Fig. \ref{figure4}b together with double exponential fits. Both the peak electronic temperature (Fig. \ref{figure4}c) and the fast time constant (Fig. \ref{figure4}d) exhibit a pronounced dip at resonance with the E$_{\text{1u}}$ mode.

Starting from a two-temperature model \cite{LinPRB2008} adapted to the present situation by introducing the electronic \cite{CastroNJP2009,KlirosArXiv2011} as well as the phonon density of states \cite{CocemasovNanoscale2015} of bilayer graphene, we find that the measured enhancement of $\lambda_{\text{e-ph}}$ is sufficient to reproduce both the reduction in peak electronic temperature and the observed decrease in cooling time (green lines in Fig. \ref{figure4}c and d, respectively). This can be understood as follows: An increase in $\lambda_{\text{e-ph}}$ decreases the Fermi velocity via $v_{\text{F}}=v_{\text{F}}^0/(\lambda_{\text{e-ph}}+1)$, where $v_{\text{F}}^0=10^6$\,m/s is the Fermi velocity in the absence of electron-phonon coupling. This results in an enhanced electronic density of states at the Fermi level and an enhanced electronic heat capacity. Thus, in the presence of an enhanced $\lambda_{\text{e-ph}}$, the same incident pump fluence will result in lower peak electronic temperatures. The decrease in cooling time is a direct consequence of the stronger coupling between electrons and optical phonons that are the major cooling channel for the hot electrons.

\section{conclusion}

The above analysis strongly indicates that the electron-phonon coupling constant is increased when the graphene lattice is modulated. Although the microscopic origin of this effect is not understood at present, we note that recent theoretical work discussed the effect of coherent modulation of the lattice and proposed a reduction of the electronic band width \cite{KnapArXiv2016}. This would explain the observed change in $\lambda_{\text{e-ph}}$. Our data may provide useful perspective for the stimulated dynamics of other carbon-based materials and possibly explain the enhancement of superconductivity observed in K$_3$C$_{60}$ \cite{MitranoNature2016}. Finally, the results also indicate that a systematic control of the electron-phonon coupling strength through phonon pumping may be possible in a broad class of materials, leading to rational design of new functional properties away from equilibrium.

\section{acknowledgment}
 
We thank A. Kemper, S. Yang, and J. A. Sobota for fruitful discussions. This work was supported by the German Research Foundation (DFG) in the framework of the Priority Program SPP1459 and the Collaborative Research Centre SFB925 as well as the European Union's Horizon 2020 research and innovation programme under grant agreement No. 696656-GrapheneCore1. Access to the Artemis facility at the Rutherford Appleton Laboratory was funded by STFC. E. P. acknowledges financial support from the Swiss National Science Foundation through an Advanced Postdoc Mobility Grant.

\clearpage

\pagebreak

\end{document}